\documentclass[conference]{IEEEtran}
\ifCLASSINFOpdf
\else
\fi

\usepackage[colorlinks, citecolor=blue]{hyperref}
\usepackage{booktabs}
\usepackage{slashbox}
\usepackage{array}
\newcolumntype{P}[1]{>{\centering\arraybackslash}p{#1}}
\usepackage{paralist}

\usepackage{setspace, amsmath,  url, lscape, subfig, multirow,   wrapfig, color, cite, bookmark} 
\usepackage[ruled,linesnumbered,inoutnumbered]{algorithm2e}
\usepackage[dvips]{graphicx}
\usepackage{epsfig}

\usepackage{xcolor}
\usepackage{balance}
\usepackage{float}

\newcommand{\paperfolder}{.}
\newcommand\Mark[1]{\textsuperscript{#1}}
\newcommand{\eg}{{\it e.g., }}

\newcommand{\ie}{{\it i.e., }}
\newcommand{\fee}{FELARE\xspace}
\newcommand{\ee}{ELARE\xspace}
\newcommand{\comments}[1]{}
\newcommand\hl{\bgroup\markoverwith
  {\textcolor{yellow}{\rule[-.5ex]{2pt}{2.5ex}}}\ULon}
  
\newlength{\boxfigwidth}

\begin{document}
\title{ \fee: Fair Scheduling of Machine Learning Tasks on Heterogeneous Edge Systems
\\[.75ex] 
  {\normalfont\large 
  Ali Mokhtari\Mark{1}, Md Abir Hossen\Mark{2}, Pooyan Jamshidi\Mark{3}, Mohsen Amini Salehi\Mark{1}%
  }\\[-1.5ex]
}

\author{
    \IEEEauthorblockA{%
        \Mark{1} \{ali.mokhtari1, amini\}@louisiana.edu\\
        HPCC Lab, School of Computing and Informatics\\
        University of Louisiana at Lafayette,
        USA
    }
    \and
    \IEEEauthorblockA{
        \Mark{2}mhossen@email.sc.edu, \Mark{3}pjamshid@cse.sc.edu  \\
        College of Engineering and Computing\\
        University of South Carolina,
        USA
    }
   
}

\maketitle              
\begin{abstract}

Edge computing enables smart IoT-based systems via concurrent and continuous execution of latency-sensitive machine learning (ML) applications. These edge-based machine learning systems are often battery-powered (i.e., energy-limited). They use heterogeneous resources with diverse computing performance (e.g., CPU, GPU, and/or FPGA) to fulfill the latency constraints of ML applications. The challenge is to allocate user requests for different ML applications on the Heterogeneous Edge Computing Systems (HEC) with respect to both the energy and latency constraints of these systems. To this end, we study and analyze resource allocation solutions that can increase the on-time task completion rate while considering the energy constraint. Importantly, we investigate edge-friendly (lightweight) multi-objective mapping heuristics that do not become biased toward a particular application type to achieve the objectives; instead, the heuristics consider ``fairness'' across the concurrent ML applications in their mapping decisions. Performance evaluations demonstrate that the proposed heuristic outperforms widely-used heuristics in heterogeneous systems in terms of the latency and energy objectives, particularly, at low to moderate request arrival rates. We observed 8.9\% improvement in on-time task completion rate and 12.6\% in energy-saving without imposing any significant overhead on the edge system. %The proposed mapping heuristic could also achieve fairness across different applications, while the energy and latency objectives are also fulfilled.

\end{abstract}


\section{Introduction} \label{sec:intro}
\subsection{Motivation}

IoT-based systems commonly rely on edge computing to process Machine Learning (ML) applications in the user's proximity, thereby, offering low-latency smart services. However, the edge systems are often battery-powered and have a limited energy supply. In many use cases, the IoT-based systems offer multiple smart services to their users (\eg object detection and motion capture \cite{bhatia2017comprehensive}). Therefore, the corresponding edge system needs to handle multiple compute-intensive ML applications simultaneously (\ie concurrently) and continuously on its limited resources. These limitations justify making use of inconsistently Heterogeneous Edge Computing (\emph{HEC}) systems \cite{salehi2016stochastic, mokhtari2020autonomous} where processors are architecturally diverse and offer different compute performance and energy consumption for distinct application types.

An exemplar use case of such an IoT-based system is SmartSight \cite{davood22} (see Figure~\ref{fig:overview}), whose aim is to offer ambient perception
to the blind and visually impaired people. The system operates based on a pair of smart-glasses and a companion edge system that takes advantage of \emph{inconsistently heterogeneous} processing units \cite{salehi2016stochastic} (a.k.a. machines) in which each machine type has a different energy consumption and is optimized for fast execution of certain task types. For instance, the HEC system can take advantage of both ESP32 (with IVP-EP 32-way SIMD imaging/video dataplane processor) \cite{9700908} and ARC EM9D/EM11D processors~\cite{EM9D} that are optimized for image/video processing tasks (\eg object detection inference), and audio processing tasks (\eg speech recognition), respectively. In this system, the companion edge system has to concurrently and continuously execute multiple ML applications and serve the user's requests within a short latency (in less than 100 milliseconds), thereby, offering services such as object detection to identify obstacles, motion detection to identify approaching objects, face recognition, text recognition, and speech recognition to respond the user's commands. Both energy-efficiency and low-latency are critical metrics for the \emph{usability} and dependability of this system and the day-to-day life of the disabled user. As such, any platform for this system must aim at \emph{maximizing the energy-efficiency} and \emph{minimizing the latency} for \emph{all} the offered services.

\subsection{Problem Statement}
\begin{figure} 
  \centering
  \includegraphics[width = 0.49\textwidth]{\paperfolder/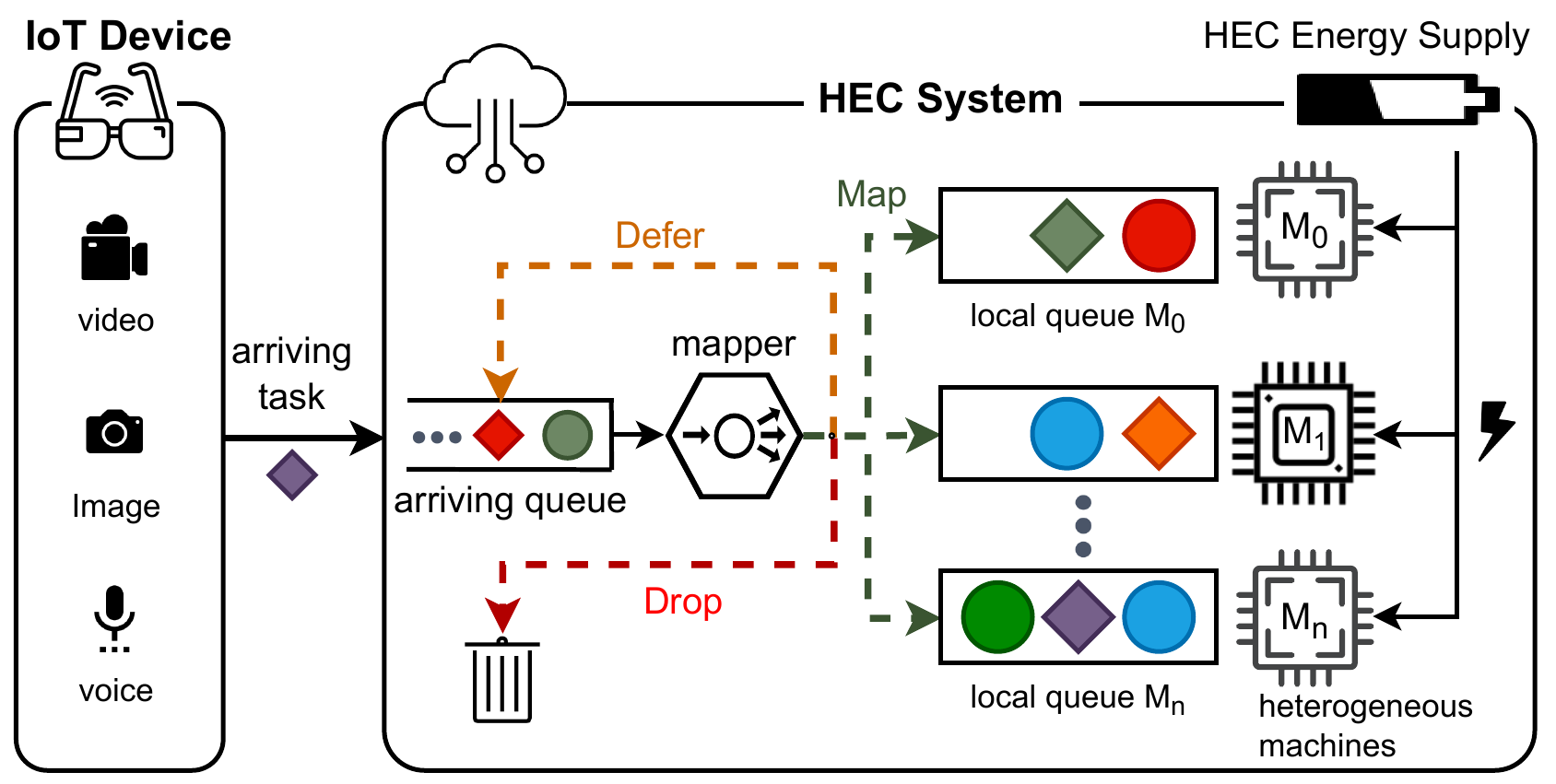}
  \vspace{-1mm}
  \caption{Overview of an Edge-based system that offers multiple ML services (\eg Object detection, speech recognition, and face recognition) concurrently using an energy- and resource-limited Heterogeneous Edge Computing (HEC) system. Sensors of the system capture multi-modal input (\eg video, image, and voice) and form latency-sensitive task requests.\label{fig:overview} }
  \vspace{-5mm}
\end{figure}

An overview of the HEC system that we consider in this study is shown in Figure~\ref{fig:overview}. Multi-modal data (\eg image, video, voice) are streamed from the sensors of an IoT device and form different types of task requests that dynamically arrive at the HEC system. Subsequently, the mapper module is triggered and allocates the tasks to a set of inconsistently heterogeneous machines. 
Before executing a task, its data is fetched to the local queue of the machine. Note that the local queues of the machines are limited. 
% Each machine owns a local queue with a limited size where the data of assigned tasks are fetched prior to their execution. 
The tasks are \emph{independent} and \emph{latency-sensitive}, with \emph{individual hard deadlines}. 

In the HEC system, efficiently allocating tasks to the limited resources is decisive on the \emph{latency} and \emph{energy} objectives. 
Intuitively, resource allocation decisions can minimize the energy consumption by mapping user requests to the machine with minimum energy usage (and computing performance). However, such allocations can potentially undermine the latency objective. In contrast, allocating user requests to the fastest machine (with higher energy consumption)  depletes the battery quickly and runs the system unusable.
In a system like SmartSight, such an unusability can potentially threaten the user's safety. These extreme cases demonstrate the importance of efficient resource allocation to increase the up-time and usability of an energy-limited HEC system.

The resource allocation decisions must also be \emph{fair} across the concurrent services. That is, the aforementioned objectives cannot be fulfilled by making the system biased towards specific task requests. For instance, to make the system energy-efficient and last for a longer time, a resource allocation method cannot consistently ignore (\ie drop) the motion detection tasks (that have long execution times) in favor of the object detection ones (that have shorter execution times). In the case of SmartSight, such a biased system makes the blind person incapable of detecting approaching objects, which again undermines its usability. As such, \emph{fairness} across request types is the third objective that has to be considered by the resource allocation of such systems.
In sum, the problem we investigate in this research can be stated as follows: \textit{how to design a fair resource allocation method for tasks (\ie requests to concurrent ML applications), such that the on-time task completion rate is maximized within the energy constraint of the HEC system?} It is noteworthy that in such a resource-limited computing system, the resource allocation method solution should be lightweight, and its incurred overhead should not worsen the system performance.

Resource allocation problem for heterogeneous systems has been widely studied with the objective of either maximizing the on-time task completion \cite{mokhtari2020autonomous } or minimizing the energy consumption \cite{gholipour2021recent}. Nonetheless, in a usable IoT-based system, both of these objectives in addition to the fairness are desired on a resource- and energy-constraint system. 

\subsection{Solution Statement and Contributions}

In addressing the multi-objective problem, we noticed that often improving the latency objective comes with the cost of energy-inefficiency and vice versa. Hence, we investigate how to efficiently strike a trade-off between the energy and latency objectives. Such a multi-objective resource allocation problem is known to be NP-complete \cite{liu2019cooper}; thus, we develop a spectrum of heuristic solutions that range from latency-aware to energy-aware ones. We note that, in the described HEC system, executing a task that is unlikely to meet its deadline wastes the available energy for its unsuccessful execution and delays the following tasks, reducing their chance of on-time completion, hence, degrading the system's usability. To mitigate this, prior to the mapping, the mapper should proactively drop tasks that are most probably unable to meet their deadline.

The performance metric for the latency objective is the number of tasks successfully that are completed on-time (\ie within the latency constraint), and for the energy objective is the amount of energy consumed to execute task requests. Then, we show that using the proposed lightweight mapping heuristic, called \emph{\ee}, we can meet both the energy and latency objectives that can dominate the widely-used resource allocation heuristics. 
In the next step, we explore the fairness aspect. For that purpose, we first formally define a measure to quantify the fairness in a HEC system. Next, we leverage the measure to propose a Fair Energy- and Latency-Aware Resource allocation on heterogeneous Edge systems, called \emph{\fee}, that can maintain fairness across task types while the energy and latency objectives are satisfied. \fee aims at minimizing the energy consumption for the mapping decisions that are expected to result in successful task completion (\ie feasible scheduling decisions). \fee also identifies the suffered task types (\ie those with the low fairness value) and dynamically adjusts its decisions to mitigate the unfairness. 

To evaluate the efficacy of the \ee and \fee heuristics, we conducted a simulation study\footnote{\url{https://github.com/hpcclab/E2C-Sim.git}} and noticed that these methods could improve the on-time task completion by 8.9\% and reduce the wasted energy due to inefficient task scheduling by 12.6\%. These heuristics could also achieve fairness while fulfilling the energy and latency objectives. In sum, the main contributions of this research are as follows:
\begin{itemize}%[wide, labelwidth=!, labelindent=5pt]
    \item Developing a measure to quantify the scheduling fairness across different task types in a computing system.
    \item Performing a multi-objective analysis of the resource allocation in the HEC system.
    \item Leveraging the multi-objective analysis to develop a fair latency- and energy-aware heuristics for concurrent ML applications in the HEC systems.
    \item Investigating the energy consumption, latency, and fairness resulted by the proposed heuristics against widely used scheduling heuristics in the HEC systems.
\end{itemize}
Although we conduct this research in the context of machine learning tasks and edge computing, our analysis and findings are not limited to this context and can be adapted to similar contexts where a set of pre-known independent task types are deployed on energy-limited heterogeneous machines.

% The rest of this paper is organized as follows:
Paper structure:
Section~\ref{sec:related} reviews prior studies. In Section~\ref{sec:sys_model}, we provide an overview of the system. Then, in Section~\ref{sec:EE}, we describe our latency- and energy-aware heuristic. In Section~\ref{sec:fairness}, we propose a measure to quantify the fairness and the fair resource allocation. Next, in Sections~\ref{sec:exp_setup} and \ref{sec:evaluation}, the performance evaluation of the methods are described. Conclusions and future works are discussed in Section~\ref{sec:conclusion}. 
\section{Related Work} \label{sec:related}
ML applications are computationally intensive, and deploying them in resource-limited HEC systems would raise two critical challenges: (i) latency and (ii) energy consumption \cite{ran2018deepdecision,huynh2017deepmon,chen2020deep}. Prior research efforts have addressed these challenges in two ways: deploying approximate computing techniques and proposing efficient resource allocation algorithms.

Approximate computing techniques can be used to improve the latency and energy consumption of the ML applications~\cite{jiang2020approximate,mrazek2016design,ansari2019improving}. Approximate computing, in particular, improves the latency and energy consumption of computationally intensive applications by leveraging their error resistance~\cite{venkataramani2015approximate}. In hardware-level approximate computing, the hardware components are modified such that the computations are performed with less energy consumption and lower accuracy. In~\cite{ansari2019improving,akbari2018px}, the researchers implemented approximate adders and multipliers in deep learning accelerators to improve the performance of the accelerator and save energy. Quantization is another approximate computing technique that can be used to reduce the latency and energy consumption of deep learning applications. 
In integer quantization, the 32-bit floating-point numbers (\eg weights) are converted to 8-bit fixed-point numbers to shrink the model, and consequently reduce the latency\cite{wu2020integer}. In~\cite{jacob2018quantization}, the authors suggested a quantization approach that allows inference to be performed using only integer arithmetic. Their proposed quantization technique could reduce the model size by 4x. In~\cite{coelho2021automatic}, the authors introduce a layer-wise and per-parameter quantization method that could maintain the accuracy while the energy consumption and model size is decreased.

The optimal scheduling is proven to be an NP-hard problem~\cite{liu2019cooper,bitam2018fog}, thereby, a substantial exploration has been accomplished to propose a feasible heuristic-based or sub-optimal solutions. The majority of the proposed solutions in scheduling focuses on one or two objectives like energy~\cite{yadav2018adaptive,zhang2018energy,ding2020q}, makespan~\cite{zhou2019minimizing}, or QoS~\cite{kumar2020pso}. In~\cite{ghanavati2020energy} a bio-inspired approach was proposed to solve the bi-objective optimization problem for the system makespan and the energy consumption objectives. In \cite{tarafdar2021energy}, the authors employed linear weighted sum techniques to minimize both energy and makespan. In~\cite{mokhtari2020autonomous,denninnart2020efficient}, the probabilistic approach is used to determine the probability of on-time completion of tasks on available computing resources. Then, a task dropping mechanism is deployed to maximize the system's robustness. The dropping decisions are made such that the overall system performance would increase.

The optimal scheduling and approximate computing techniques are complementary solutions. In fact, the synergy between approximate computing and efficient scheduling can significantly increase the overall system performance. In this work, we propose an efficient resource allocation algorithm that aims to minimize the system's wasted energy due to unsuccessful completion of tasks while the performance metric is maintained. Furthermore, the proposed method is developed first to detect the unfairness across task types, and then it treats the suffered task types until the fairness criteria are satisfied.  
\section{System Model } \label{sec:sys_model}
This research encompasses scenarios where a single-user HEC system is employed to host multiple ML applications, such as SmartSight~\cite{davood22} and those explored in \cite{aazam2021task,kherraf2019optimized,rahman2021iot}. In these scenarios, ML applications are latency-sensitive and have to process the user's requests in a real-time manner. As an example, a request to an obstacle detection application in SmartSight has to detect the objects (obstacles) for a disabled user within a short hard deadline. There is no use in completing a task after the deadline has passed, and doing so makes the solution unusable for the disabled user. Moreover, these HEC systems host only a \emph{limited} and pre-known ML applications (a.k.a. task types) with different data modalities (\eg video, image, or voice) that are assumed to have the same priority. As shown in Figure~\ref{fig:overview}, task with various modalities are queued upon arrival. Then, the resource allocator (a.k.a. scheduler) uses a mapping heuristic to make one of the following decisions for each queued task: (A) \emph{mapping} it to an available slot in the local queue of a machine; (B) discarding it, via \emph{dropping} it or \emph{deferring} its mapping to a later time. A mapping could happen in two situations: (i) completion of an executing task or (ii) arrival of a new task. The local queue on each machine is to fetch the required data (\eg image, audio, or video) and prepare the assigned tasks for execution.  Due to uncertainty in the execution time of tasks and its compound impact on the mapping decisions, the local queues are considered to have a limited size, and they are equal across different machines in the system \cite{ipdps19,chavithcw19}. Furthermore, task requests dynamically arrive to the system and the order of arrival is unknown. 

%We benchmarked several ML applications and noticed that the execution times of the task requests for the same ML application have insignificant variations. The reason is even though the input data are different, the task requests use the same ML model, thus, have the same computing requirements and exhibit a similar execution time on each machine type. As such, 
We categorize the arriving tasks based on the ML application they belong to, and call them \emph{task types}. Similarly, heterogeneous machines in the HEC systems are distinguished by their performance characteristics and architectures and are considered as different \emph{machine types}. 
Profiling the execution time of task types  on the machines provides information about the execution time of task type $i$ on machine type $j$. Then, the expected values of the execution times for all task types on different machine types are utilized to form a matrix, called \emph{Expected Execution Time (EET)} matrix \cite{salehi2016stochastic}. The  number of task types and the number of machine types in the HEC system determine the number of rows and columns of the EET matrix. In this work, we assume that the EET matrix is available via leveraging task profiling data of the HEC system.

To determine the energy consumption of the system, we use the idle and dynamic powers of each machine. Specifically, the amount of the energy used by a machine of type $j$ to process a task of type $i$ is determined by multiplying the expected execution time of task type $i$ by the dynamic power of that machine $j$. For an idle machine, the amount of the energy used is determined by multiplying the idle time by its idle power. Due to the data transfer overhead and latency constraint of the tasks, we assume that a mapped task cannot be remapped or preempted. Machine queues are also served in a first come, first served manner. 

\section{\ee: Energy- and Latency-aware Resource Allocation in Heterogeneous Edge Computing} \label{sec:EE}

\subsection{Overview}

In the HEC systems, the energy consumption due to processing a latency-sensitive task is wasted in two ways: (i) unsuccessful task completion; and (ii) inefficient resource allocation. The former is caused by mapping a task to a machine where the task misses its deadline (\ie the expected completion time of the task on that machine is greater than its deadline). The latter explains the extra energy consumed (wasted) by a machine to successfully complete a task that could be otherwise successfully completed on another machine with less consumed energy.

To mitigate the energy wastage, we propose a two-phase Latency- and Energy-aware resource allocation method, called \emph{\ee}, to map tasks in the HEC system. To tackle the wasted energy due to unsuccessful task completion, in the first phase, the mapper identifies the unlikely-to-succeed tasks in the arriving queue and defers their assignment to the next mapping events with the hope that a better matching machine would be available at that time. However, it is possible that the task deferral continues in the following mapping events until the task's deadline is violated. In this case, the task is dropped and the system would not further process that task. To mitigate the wasted energy due to inefficient resource allocation, the mapper minimizes the energy consumption in its mapping decisions for the feasible tasks in the arriving queue. Algorithm~\ref{alg:EE} provides the pseudo-code of the \ee heuristic. In the following sections, we elaborate on the details of each phase of this heuristic.

\begin{algorithm}[h]
    \caption{\ee Heuristic}
 	\label{alg:EE}
	\SetAlgoVlined
	\SetKwFor{ForEach}{for each}{do}{end}
 	\DontPrintSemicolon
	\SetKwBlock{Function}{Function \texttt{ EE(ArrivingQueue, Machines, EET)}}{end}

	\KwIn{ArrivingQueue: $\{T_0,\ T_1,\ \cdots,\ T_{p}\}$ \;
	       Machines: $\{M_0,\ M_1,\ \cdots,\ M_m\}$\;
	       EET: Expected Execution Time Matrix\;}
	\KwOut{Mapping Pairs: A list of [task, machine] \; }
	
 	\Function{
	  Call Phase-I with the ArrivingQueue, Machines \& EET as inputs to generate feasible task-machine pairs,\;
	  and the set of infeasible tasks\;
	  \ForEach {task $\in$ infeasible tasks} {
	  
	    \eIf {task.deadline $<$ current time } {
	            defer(task)\;
	        }
	        {
	            drop(task)\;
	        }

	   }
	  Call Phase-II with feasible task-machine pairs \& Machines as inputs \;
	  \KwRet set of task-machine pairs for mapping  \;
	}
\end{algorithm}

\subsection{Phase-I: Latency- and Energy-Awareness }
Recall that the Phase-I of \ee is responsible for recognizing the infeasible tasks in the arriving queue. A $[task, machine]$ pair is deemed as \emph{feasible} pair if machine can successfully complete the task. A task that appears in at least one feasible $[task, machine]$ pair is identified as a feasible task. To determine the feasibility of a $[task,machine]$ pair, the expected completion time of the task on the machine is required.
Let task~$i$ with deadline $\delta_{i}$ map to slot~$q$ of machine~$j$ and is given a start time, denoted by $s_{ij^{(q)}}$, when the machine is being idle. The expected execution time of processing task~$i$ on machine~$j$ is $e_{ij}$ time units, which is extracted from the EET matrix. Thus, the expected completion time of task~$i$ when it is mapped to the queue slot~$q$ of machine~$j$, denoted by $c_{ij^{(q)}}$, is calculated based on  Equation~\ref{eq:compl}: 
\begin{equation}\label{eq:compl}
       c_{ij^{(q)}}= 
    \begin{cases}
            s_{ij^{(q)}} + e_{ij}, &   s_{ij^{(q)}} + e_{ij} < \delta_i \\
       
            \delta_i, &   s_{ij^{(q)}} + e_{ij} > \delta_i  \ and \  s_{ij^{(q)}} < \delta_i\\
        
            s_{ij^{(q)}} , &   s_{ij^{(q)}} \geq \delta_i
    \end{cases}
\end{equation}

In Equation~\ref{eq:compl}, the first row belongs to feasible pairs while two other rows belong to infeasible pairs. In case of missing the deadline, the completion time of the task could be either its deadline (\ie task is dropped during execution immediately when it passes its deadline) or the start time (\ie the task is dropped before execution started because the task has already passed its deadline). 

In Phase-I, to prevent inefficient scheduling, for each task in arriving queue, the feasible $[task, machine]$ pair that incurs minimum energy usage is singled out and considered as the feasible and efficient pair for that task. To that end, the expected energy consumption for executing task~$i$ when it is mapped to queue slot~$q$ of machine~$j$ is determined as follows:
\begin{equation}\label{eq:engy}
    ec_{ij} = 
    \begin{cases}
           
            p^{dyn}_j \cdot \ (\delta_i - s_{ij^{(q)}}) , &   c_{ij^{(q)}} > \delta_i  \ and \  s_{ij^{(q)}} < \delta_i\\
             p^{dyn}_j \cdot \ e_{ij}, &   c_{ij^{(q)}} < \delta_i \\
            0 , &   s_{ij^{(q)}} \geq \delta_i
    \end{cases}
\end{equation}

In Equation~\ref{eq:engy}, the first row describes the wasted energy due to unsuccessful completion of the task. In the case of successful task completion, the middle term of the Equation~\ref{eq:engy} gives the amount of energy consumed by the machine. However, this energy consumption is not always the optimal value for completing the task. In other words, a scheduler that makes an inefficient scheduling decision would result in higher energy consumption, thus, wasting energy. 
In the case that a feasible task appears in multiple $[task, machine]$ pairs, all the pairs except the one with the minimum energy consumption are considered as ``inefficient pairs'' that acting upon them increase the energy consumption, thus, reducing the usability of the HEC system. To avoid this energy wastage, for each task, its feasible pair with minimum expected energy consumption (efficient feasible pair) is selected in Phase-I.

The pseudo-code for Phase-I is shown in Algorithm~\ref{alg:Ph1}. Lines 6-11 generate a list of feasible machines for each task in the arriving queue and their corresponding Expected Energy Consumption. In Line 13, the feasible machine with minimum expected energy consumption (EEC), $ec_{ij}$, is selected. Then, in Line 14, task $T_i$ and its matching machine with minimum EEC (\ie efficient and feasible machine for completing the task) create a pair, and it is appended to the list of feasible efficient pairs. In Line 16, a task that has no chance of being completed by any available machines is considered an infeasible task and stored in the list with the same name.

\begin{algorithm}
	\caption{Phase-I of \ee Heuristic}
 	\label{alg:Ph1}
	\SetAlgoVlined
	\SetKwFor{ForEach}{for each}{do}{end}
 	\DontPrintSemicolon
	\SetKwBlock{Function}{Function \texttt{Phase-I(ArrivingQueue, Machines, EET)}}{end}

	\KwIn{ArrivingQueue: $\{T_0,\ T_1,\ \cdots,\ T_{p}\}$\;
	 Machines: $\{M_0,\ M_1,\ \cdots,\ M_n \}$\;
	 EET: Expected Execution Time Matrix\;}
	
	\KwOut{Feasible Efficient Pairs, Infeasible Tasks\;}
	
	\Function{
	 
	  \ForEach{$T_i \in ArrivingQueue $}{
	    
    	    \ForEach{$M_j \in Machines$}{
    	     Calculate $c_{ij}$ using Equation~\ref{eq:compl} \;
    	    \If{ $c_{ij} \leq \delta_i$ }{
    	       Calculate $ec_{ij}$ using Equation~\ref{eq:engy} \;
    	       Append $[M_j, ec_{ij}]$ to FeasibleMachines\;}
    	    }
    	    
    	    \eIf{$FeasibleMachines \neq NULL$}{
        	    $M_{eff}$ = Select machine with min $ec_{ij}$ in FeasibleMachines \;
    	        Append [$T_i$ , $M_{eff}$] to Feasible Efficient Pairs \\ 
    	   }{
    	        Apppend $T_i$ to Infeasible Tasks\;
    	   }
	   }  
	\KwRet  Feasible Efficient Pairs, InfeasibleTasks \;
	}
\end{algorithm}

\subsection{Phase--II: Minimizing the Energy Consumption}
Recall that the output of the Phase-I is the set of feasible and efficient $[Task, Machine]$ pairs. It is possible that multiple tasks become feasible for for a machine in the system. To avoid wasting energy due to inefficient resource allocation, Phase-II of the \ee heuristic is responsible to map the feasible pair that incurs the minimum expected energy consumption. 

\begin{algorithm}
	\SetAlgoVlined
	\DontPrintSemicolon
	\SetKwFor{ForEach}{for each}{do}{end}
	\SetKwBlock{Function}{Function \texttt{ Phase-II(Machines, Feasible Efficient Pairs)}}{end}

	\KwIn{Machines: $\{M_0,\ M_1,\ \cdots,\ M_n \}$,\;
	      Feasible Efficient Pairs: A list of $[T_i,M_j, EEC_{ij}]$\;}
	\KwOut{ Mapping Pairs: A list of [task, machine]\; }
	
	\Function{
	  
	  \ForEach{$M_j \in Machines$ }{
	        
    	   $Nominees_j \leftarrow$ Call Get\_Nominees(Feasible Efficient Pairs, $M_j$)\;
    	    
    	    $T_i \leftarrow$ Select the task with min $ec_{ij}$ from $Nominees_j$\;
    	    Append $[T_i,M_j]$ to the list of Mapping Pairs\;
	       
	   }
	   \KwRet Mapping Pairs \;
	   
	}
	\caption{Phase-II of the \ee Heuristic}
	\label{alg:Ph2}
	
\end{algorithm}%%%%%

Algorithm~\ref{alg:Ph2} shows the pseudo-code for the Phase-II of the \ee heuristic.
In Line 6, all tasks that match Machine $j$ are retrieved. Then, in Line 7, the most efficient task (\ie with minimum energy consumption) is selected, and appended to the mapping pairs in Line 8. Next, the mapping pairs are returned to the main \ee heuristic.
\section{Fairness in Completing Task Types} \label{sec:fairness}

A resource allocation method is deemed fair if it is unbiased in allocating resources to the tasks of the same priority. That is, the resource allocation should not prioritize task types based on their execution time or any other system level metric other than those explicitly defined by the user. Recall that we assumed no precedence across task types in HEC. Therefore, we can use the successful completion rate of different task types as the metric to measure the fairness across all task types. The completion rate of task type $i$, denoted $cr_i$, represents the portion of the tasks of type $i$ that are successfully completed within a given time interval. In other words, task type completion rate is the ratio of the number of completed tasks on-time for a certain task type to the total number of tasks of that type arrived to the system. In an ideal and fair resource allocation, the completion rate of all task types is one ($\forall{i}$ $cr_i = 1 $).
However, due to non-optimal resource allocation or shortage in resources, some tasks may not meet their deadlines, thus, their task type completion rate decreases by each missing task. To quantify the fairness, we continuously monitor the task types completion rates. An observed completion rate distribution could lie in one of the following categories: (i) Co-existence of high and low values for task type completion rates; (ii) Similar but low completion rate values for all task types; or (iii) Similar and high completion rate for all the task types. The first observation describes the situation where the mapping method favors certain task types (with high completion rates) over others (with low completion rates). In contrast, the third one represents a mapping heuristic that exhibits fairness across all the task types. Accordingly, improving the fairness in a biased mapping system is translated as moving from category (i) to (iii).

\begin{algorithm}
	\SetAlgoVlined
	\DontPrintSemicolon
	\SetKwFor{ForEach}{for each}{do}{end}
	\SetKwBlock{Function}{Function \texttt{Suffered(TaskTypes)}}{end}
	
	\KwIn{TaskTypes: $\{TT_0,\ TT_2, \ \cdots,\ TT_s \}$\;}
	
	\KwOut{Suffered Task Types \; }
	
	\Function{
	    
	   Calculate $\mu$ and $\sigma$ of the task type completion rates\;
	   Calculate fairness limit, $\epsilon$ using Equation~\ref{eq:fairrange} \;
	   
	  \ForEach{$TT_i  \in TaskTypes $ }{
	       $cr_i$: completion rate of $TT_i$ \;
    	    \If{ $cr_i \leq \epsilon$ }{
    	        Append $TT_i$ to Suffered Task Types\;}
	   }  
	\KwRet  Suffered Task Types \;

	}
	\caption{Suffered Task Types}
	\label{alg:suffered}
	
\end{algorithm}%%%%%

\begin{figure*}
     \centering
     \includegraphics[width=0.85\linewidth, height=0.24\linewidth]{\paperfolder/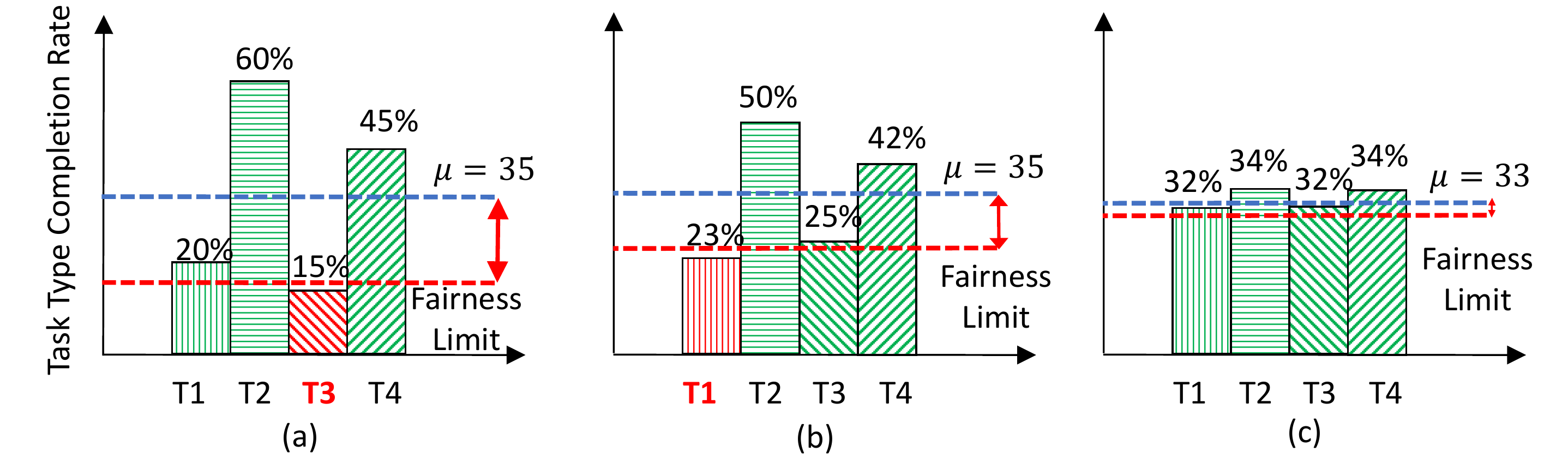}
     \vspace{-3mm}
     \caption{Illustration of the fairness limit method at different mapping events. The completion rate variance is diminishing from left to right, as a result of \fee mapping heuristic. Suffered task types (T1 and T3) are gradually treated by fee heuristic.}
      \vspace{-3mm}

  \label{fig:fairness_limit}
\end{figure*}
 
In the case of co-existing high and low completion rates, the task types with low completion rates are the \emph{suffered task types}. To identify the suffered task types, we define a \emph{fairness limit}, denoted $\epsilon$, such that any task type whose completion rate is lower than this limit is considered as suffered task type. To calculate this limit, we introduce the \emph{fairness factor}, denoted $f$, that represents the aggressiveness of the fairness method. Let $\mu$ and $\sigma$, respectively, represent the mean and standard deviation of the completion rates across all task types. Then, the the fairness limit is calculated based on Equation~\ref{eq:fairrange}. A higher value for the fairness factor results in a less aggressive fairness method. In the extreme case where $f$ is enough large, the fairness limit approaches zero, thus, does not identify any suffered task types (\ie the fairness method is disabled). 
\begin{equation}\label{eq:fairrange}
    \epsilon = \mu - f\cdot\sigma \quad \textrm{where}  \ 0 \leq f \leq \frac{\mu}{\sigma}
\end{equation}

Upon calculating the fairness limit, any task type whose completion rate is below the limit is identified as a member in the set of suffered task types, \ie $cr_i < \epsilon \Longleftrightarrow i \in \textrm{\emph{ suffered task types}}$. Figure~\ref{fig:fairness_limit} further clarifies the way to identify the suffered task types using the fairness limit method. In subfigure (a), the completion rate for task types $\{T_1, T_2, T_3, T_4\}$ is 20\%, 60\%, 15\%, and 45\%, respectively. To identify the suffered task types, the mean and standard deviation of the completion rates across all task types is calculated; we have $\mu=35$ and $\sigma=18.4$. Assuming the fairness factor be  one (\ie $f=1$), based on Equation~\ref{eq:fairrange}, the value of the fairness limit becomes $\epsilon=16.6$. Because completion rate of $T_3$ is less than the fairness limit ($cr_3=15\%$), it is identified as suffered task type. In next mapping events, the \fee method gives $T_3$ a higher priority, thus, its completion rate increases ($cr_3 = 25\%$). From subfigure (b) to (b), Although the mean of the completion rates does not change ($\mu = 35$), the standard deviation ($\sigma = 11.4$) decreases as a result of applying \fee mapping heuristic. Therefore, the fairness limit is shrinking and $T_1$ is identified as a suffered task type ($cr_1 = 23 < 23.6 $), thus, the higher priority is given to $T_1$ in next mapping events. Eventually, the standard deviation converges to zero and the the gap between mean and fairness limit diminishes, as depicted in subfigure (c). 

Algorithm~\ref{alg:suffered} explains the pseudo-code to identify suffered task types. In Line 4, the mean and standard deviation of the task type completion rates are calculated. Then, in Line 5, the Equation~\ref{eq:fairrange} is used to determine the fairness limit. Eventually, in Lines 6-9, the completion rate of each task type, $cr_i$ is compared against the fairness limit to identify the suffered task types.

Once we know the suffered task types, we leverage it to make the system fair. For that purpose, we extend the \ee heuristic and propose a new heuristic, called Fair Energy- and Latency-aware Resource Allocation (\fee). It follows the following approaches to address the fairness:
\begin{itemize}
    \item Prioritizing the suffered tasks in the mapping events.
    \item Leveraging task dropping for non-suffered tasks in favor of infeasible suffered tasks to make them feasible.
\end{itemize}

Using \fee, in each mapping event, the suffered task types are prioritized in allocation. Moreover, for a suffered task that is infeasible, the pending tasks in the local queue of the fastest (\ie best-matching) machine are dropped one-at-a-time, until the suffered task becomes feasible on that machine. These two strategies ultimately enhance the completion rate for the suffered tasks and gradually diminishes the dispersion in the completion rates of the task types. It is noteworthy that, once the completion rate of task type $i$ becomes greater than the fairness limit, it is removed from the list of suffered task types. To prioritize the suffered task types in mapping events, we introduce \emph{high-priority pairs} that include the feasible efficient pairs of suffered task types. To construct the high-priority pairs, we first generate the feasible efficient pairs as explained in Phase-I of \ee. Next, each pair with a task not identified as suffered is removed from the list. The resultant list contains high-priority pairs. 

% In fact, the high-priority pairs are the feasible efficient pairs generated in Phase-I of \ee, excluding not suffered task types. 
% Although prioritizing the suffered task types improves the fairness, it can degrade the overall system's performance by lowering the mean value of $\alpha$. To address this problem, we consider a \emph{prioritization chance}, denoted by $\beta$, that indicates the probability of prioritization in mapping events. Therefore, we randomly (with the probability of $\beta$) prioritize the suffered task types by replacing the feasible efficient pairs with the high-priority pairs. 
The high-priority pairs are passed to Phase-II of \fee, which is the same as Phase-II of \ee,  to make the mapping decisions.

\section{Experimental Setup} \label{sec:exp_setup}

\subsection{overview} 
To evaluate the performance of the proposed heuristics, we examine two scenarios: (i) using two deep learning applications (namely, face recognition \cite{schroff2015facenet} and speech recognition \cite{amodei2016deep}) as the task types running on two AWS Virtual Machines (VMs) (t2.xlarge and g3s.xlarge)\footnote{\url{https://aws.amazon.com/ec2/instance-types/\#Intel}}) as the machine types; and (ii) simulating the HEC system with four machine types and four task types with synthesized expected execution time (EET) matrix. 

In the first scenario, %we run two deep learning applications (\ie face recognition and speech recognition) on two EC2 instances of AWS (t2.xlarge and g3s.xlarge)\footnote{https://aws.amazon.com/ec2/instance-types/#Intel}.
the \texttt{T2} instances of AWS are general-purpose machines that can be used for various workloads. They utilize Intel Xeon processors (Haswell E5-2676 v3 or Broadwell E5-2686 v4). The \texttt{t2.xlarge} instance has 4 vCPUs and 16 GB memory. The \texttt{G3} instances are best-matched for applications with intensive graphic and equipped with NVIDIA Tesla M60 GPUs. The Thermal Design Power (TDP) of Haswell E5-2676 v3 and NVIDIA Tesla M60 is 120 W and 300W, respectively.

To perform the face recognition, we first use MTCNN model~\cite{zhang2016joint} to detect the faces, then FaceNet model~\cite{schroff2015facenet} is used to extract the embeddings. Eventually, Support Vector Machine (SVM) is used to classify the face embeddings. The input images consist of 30 images sampled from the LFW dataset~\cite{LFWTechUpdate}. Each AWS instance is used to process all input images. In addition, the experiment is repeated 30 times, and eventually we collected the end-to-end latency for 900 inferences. For speech recognition, we used the DeepSpeech model~\cite{amodei2016deep}. The test dataset consists of 900 recorded audio that sums up to 118.9 hours of speech. Lastly, we use the collected execution times to construct the EET matrix. 

To study the mapping heuristics behavior, we simulate a HEC system with four machine types and four task types. The four machines (\{$m_1$, $m_2$, $m_3$, $m_4$\}) have dynamic power consumption of $\{1.6\cdot p, 3.0\cdot p, 1.8\cdot p, 1.5\cdot p\}$ and idle powers of $0.05\cdot p$ where $p$ represents the unit power. The four heterogeneous task types are $\{T_1, T_2, T_3, T_4\}$. To model the heterogeneity of the HEC systems, we use the Coefficient-of-Variation-Based (CVB) technique\cite{ali2000representing} to populate the Expected Execution Time (EET) matrix. In the CVB method, the Coefficient of Variation (CV) of execution time values is used to measure the heterogeneity. Then, based on the task and machine CVs, two Gamma distributions are utilized to generate the expected execution times. The EET matrix is shown in Table~\ref{table:eet}. Next, the expected values in EET matrix are used to sample the execution time for each individual task from a Gamma distribution. For each task type, we calculate the average of expected execution times on machine types, denoted $\bar{e_i}$. Then, we take the average of $\bar{e_i}$ for all task types, denoted $\bar{e}$, as the average of collective expected execution time of all task types on all machine types. Finally, we determine the deadline of the task $k$ of type $i$, $\delta_i(k)$, that arrives at $arr_k$ by adding task type and collective mean values to the task's arrival time, as shown in Equation~\ref{eq: deadline}.
% the deadline of task $i$, $\delta_{i}$, is generated by summing the arrival time of the task, denoted $arr_i$, with the average execution time of the task type on all machines, denoted $\bar{e_i}$, and the average execution time of all task types on all machines, denoted $\bar{e}$. 
\begin{equation} \label{eq: deadline}
    \delta_i(k) = arr_k + \bar{e_i} + \bar{e}
\end{equation} 

We assume the inter-arrival between tasks follows the Poisson distribution \cite{praveen2018effective}.
% Finally, we consider that tasks arrive dynamically according to a Bernoulli process with certain average arrival rate.  

\begin{table*}[h]
\begin{center}
\begin{tabular}{|c|c|c|c|c|}\hline
\backslashbox{Tasks}{Machines}
&\makebox[2em]{$m_1$}&\makebox[2em]{$m_2$}&\makebox[2em]{$m_3$}
&\makebox[2em]{$m_4$}\\\hline
$T_1$ &2.238&1.696&4.359&0.736\\\hline
$T_2$ &2.256&1.828&4.377&0.868\\\hline
$T_3$ &2.076&1.531&5.096&0.865\\\hline
$T_4$ &2.092&1.622&4.388&0.913\\\hline
\end{tabular}

% \vspace{0in}
\caption{Expected Execution Time (EET) matrix. Each entry $(i,j)$ represents the expected execution time of task type $i$ on machine type $j$. The CVB technique\cite{ali2000representing} was used to generate the EET matrix.}
\label{table:eet}
\end{center}
\vspace{-6mm}
\end{table*}

\subsection{Baseline Mapping Heuristics}
Two-phase mapping heuristics have been extensively studied in heterogeneous systems~\cite{salehi2016stochastic, shi2019performance}. Here, we focus on Minimum Completion Time-Minimum Completion Time (MM), Minimum Completion Time-Soonest Deadline (MSD), and Minimum Completion Time-Maximum Urgency (MMU) as baseline heuristics. The first phase of MM, MSD, and MMU are similar. The mapper selects a [task, machine] pair with a minimum expected completion time in the first phase. Then, the list of [task, machine] pairs are used in the second phase to select an individual task for each of the available machines. The methods are distinguished based on their second phase algorithm. In MM, in the second phase, if there exist multiple tasks for a machine, the [task, machine] pair that offers minimum expected completion time is chosen, and then the task is allocated to the local queue of that machine. In MSD, it chooses the pair based on the soonest deadline. That is, for each available machine, it explores all the pairs generated in the first phase, and then it chooses the task with earliest deadline. In case of same deadline for multiple tasks, the task with minimum expected completion time is selected and assigned to queue of the machine. In MMU, 
it uses the urgency metric for selecting a task. The urgency of task $k$ of type $i$ is defined as $1 / (\delta_{i}(k) - e_{ij})$. So, the task with maximum urgency is selected and assigned to the local queue of the machine.
\begin{figure}[b]
     \centering
     \includegraphics[width=0.8\linewidth]{\paperfolder/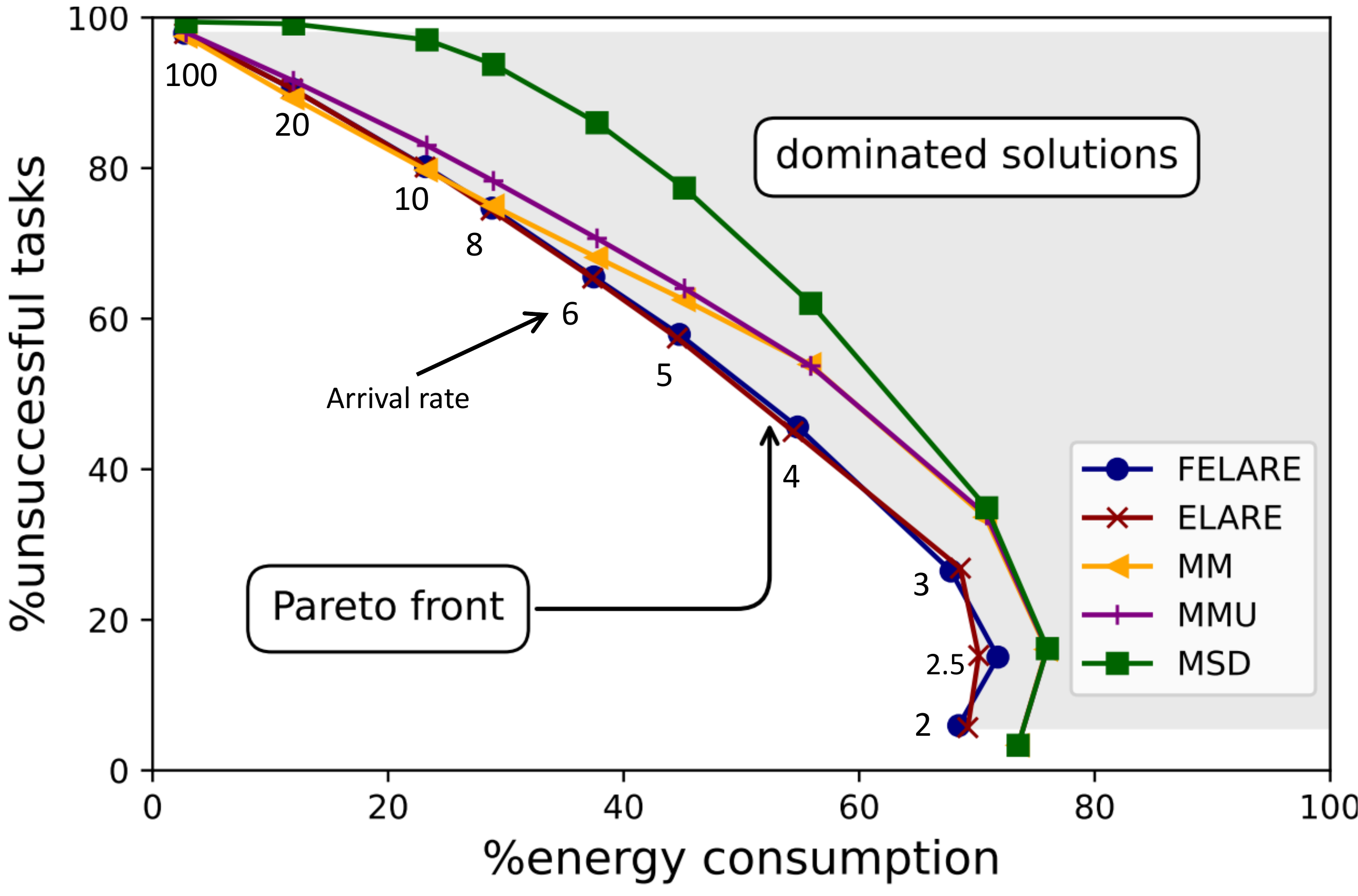}
     \vspace{-1mm}
     \caption{The trade-off between energy consumption and latency across ELARE, FELARE, and other baseline heuristics. Each curve belongs to one mapping heuristic at different arrival rates. The gray zone describes the solutione that were dominated by the Pareto-front curve. Both ELARE and FELARE are non-dominated solutions that form the Pareto-front. }
      \vspace{-3mm}

  \label{fig:pareto}
\end{figure}

\section{Performance Evaluation}\label{sec:evaluation}

\subsection{Energy and Latency Trade-off}
Recall that energy- and latency make the resource allocation problem of HEC systems a bi-objective optimization problem. Minimizing the energy consumption is conflicting with maximizing the completion rate and/or minimizing the deadline miss rate of tasks. As a result, there is not a single optimal solution, instead, there could be a set of solutions that dominate other solutions. Figure~\ref{fig:pareto} shows the energy consumption and deadline miss rate for the mapping heuristics at different arrival rates. Moving from right to left on each curve increases the arrival rate. We can observe that at extremely high arrival rate (\eg 100 tasks per second), all methods exhibits similar performance (high miss rate with low energy consumption). In fact, at very high arrival rate, a resource-limited system is highly oversubscribed and tasks are missed regardless of the applied mapping heuristic. However, the proposed heuristics, \ee and \fee, dominate other heuristics at lower arrival rates. In other words, we can say that \ee and \fee at low to moderate arrival rates belong to the set of non-dominated solutions or the Pareto-front. This analysis recommends us to employ \ee and \fee, particularly, at low to moderate arrival rates.

\subsection{Analyzing the Wasted Energy}
This experiment is to examine the performance of \ee and \fee on minimizing the energy wasted due processing infeasible tasks. To this end, we used 30 synthesized workload traces with different arrival rates where each workload trace included 2,000 tasks. We also measure the wasted energy as the percentage of energy consumed by machines to process the missed tasks with respect to the initial available energy of the HEC system. 

\begin{figure}
     \centering
     \includegraphics[width=0.8\linewidth]{\paperfolder/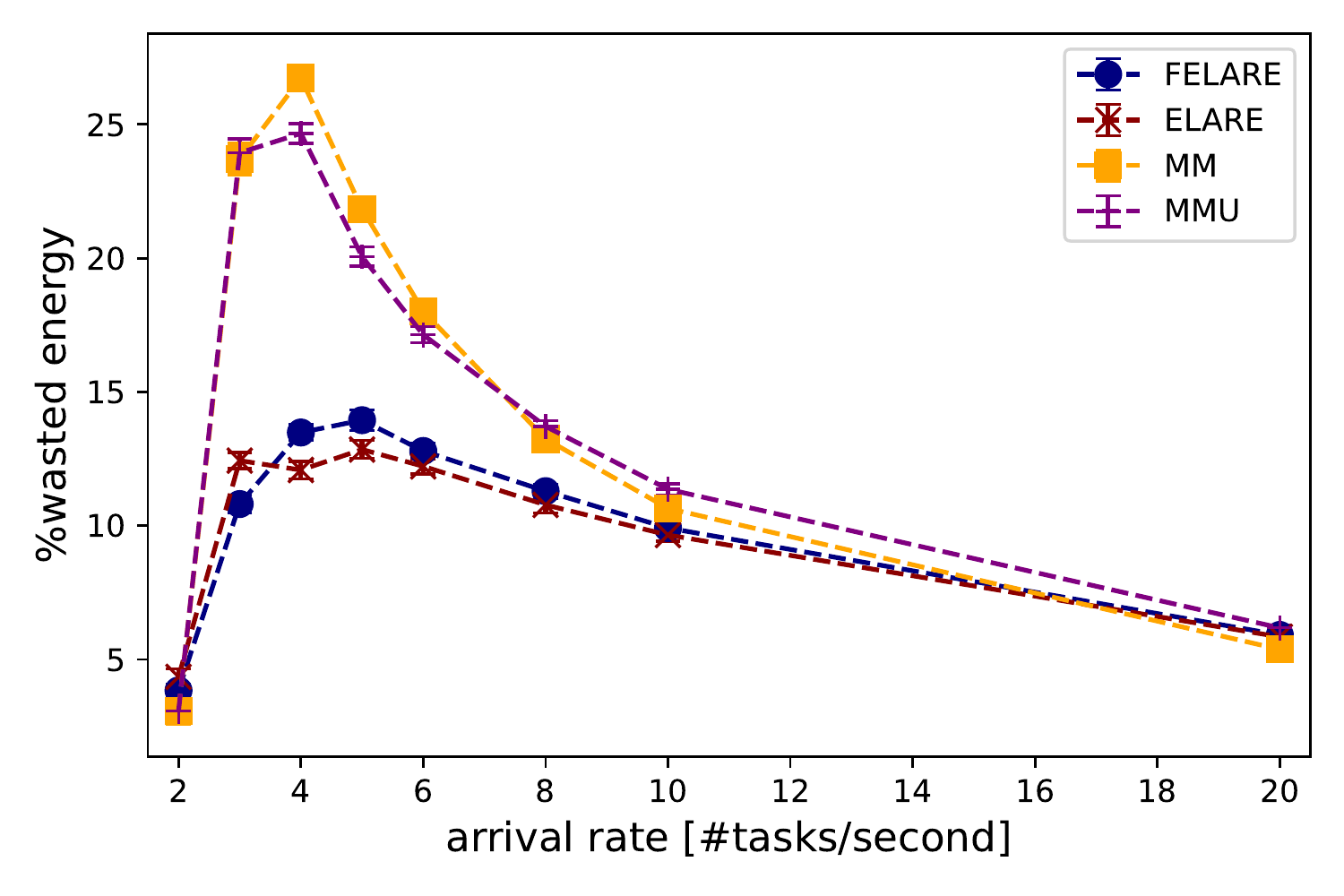}
     \vspace{-2mm}
     \caption{The wasted energy due to deadline miss at different arrival rates for different mapping heuristics}
      \vspace{-6mm}
  \label{fig:wasted}
\end{figure}

\begin{figure}[h]
     \centering
     \includegraphics[width=0.8\linewidth]{\paperfolder/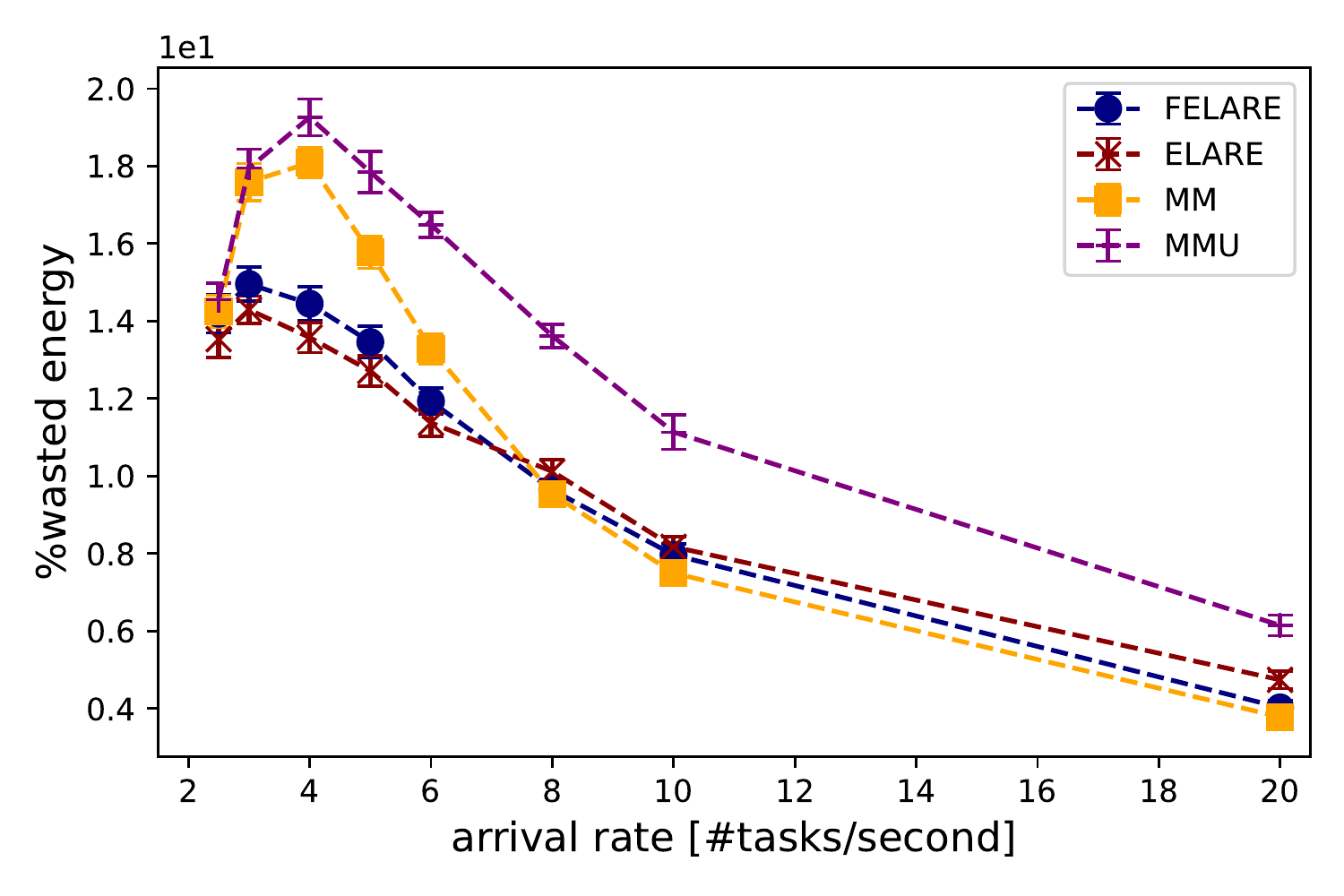}
     \vspace{-3mm}
     \caption{The wasted energy due to deadline miss of face recognition and speech recognition applications on AWS instances at different arrival rate for MM and EE }
      \vspace{-4mm}
  \label{fig:wasted_real}
\end{figure}

Figure~\ref{fig:wasted} shows the results of the wasted energy analysis. We observe that the wasted energy for \ee and \fee at low to moderate arrival rates is much less than the other heuristics. Specifically, deploying \ee shows 12.6\% less wasted energy at arrival rate of 4 tasks per second than the MM heuristic. However, for the high arrival rates, all the heuristics converge to a low energy wastage. The reason is that, at the high arrival rates, most of the tasks become infeasible and there is no chance to make them feasible. Hence, they miss their deadline before even being assigned, regardless of the mapping heuristic being used. Similar trend is observed in Figure~\ref{fig:wasted_real} where the deep learning applications are executed on the AWS instances. 

These observations can be explained by considering how the energy is consumed in each one of the heuristics. We explore this granular behavior of heuristics in the next paragraph. However, the observed results confirm that \ee and \fee methods definitely waste less energy than others for the low to medium arrival rates.

\begin{figure}
     \centering
     \includegraphics[width=0.8\linewidth]{\paperfolder/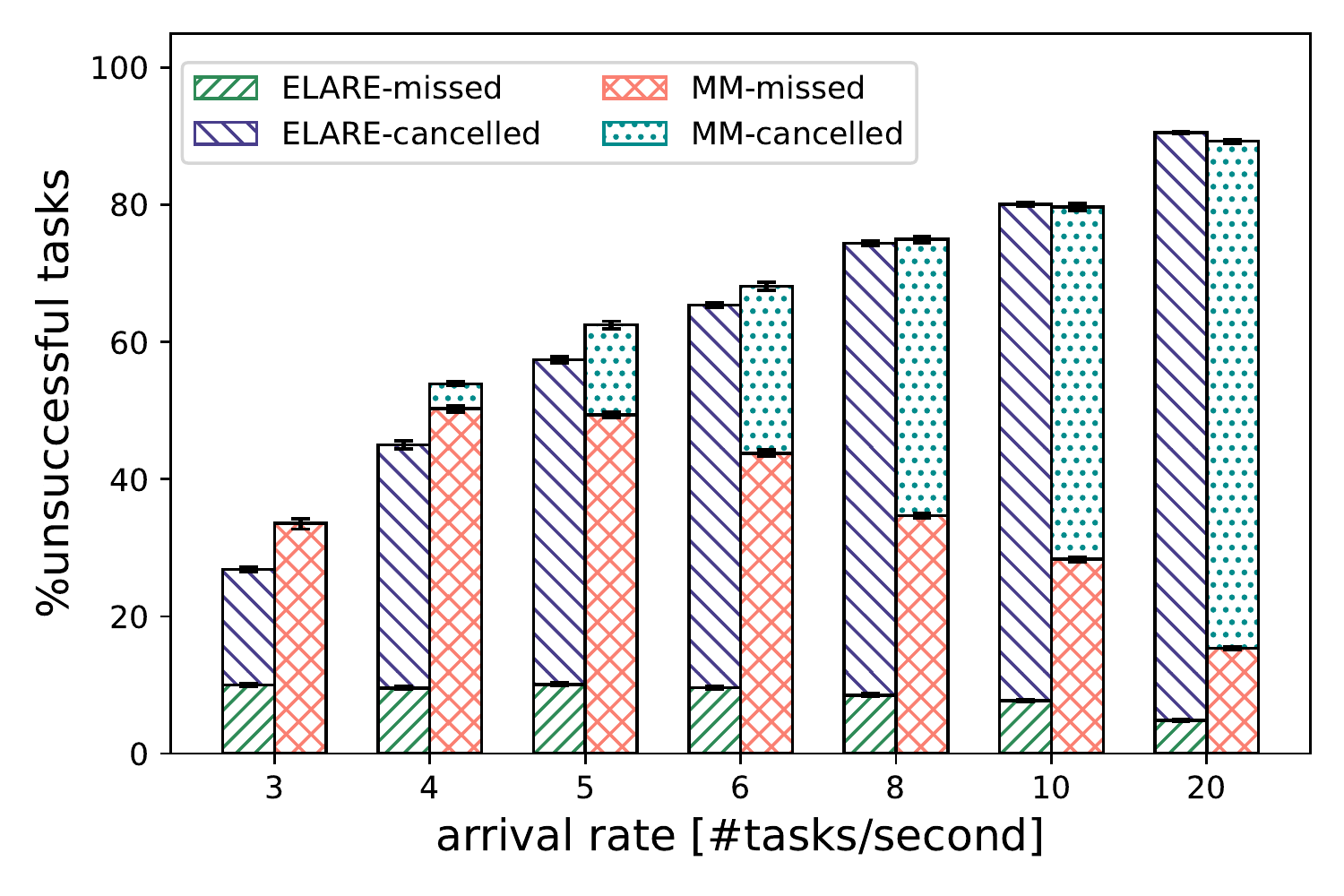}
     \vspace{-3mm}
     \caption{The percentage of unsuccessful tasks at different arrival rates for MM and \ee. Unsuccessful tasks are those that either cancelled (not assigned to the machines) or dropped due to missing deadline.}
      \vspace{-4mm}

  \label{fig:missed_cancelled}
\end{figure}

Figure~\ref{fig:missed_cancelled} shows the percentage of unsuccessful tasks, due to missing deadline or cancellation (dropping) before the assignment, for both MM and \ee at different arrival rates. We observe that \ee outperforms MM for the lower arrival rates. Specifically, \ee reduces the unsuccessful tasks by 8.9\% for the arrival rate of 3. We see that \ee proactively cancels most of the unsuccessful tasks, whereas, majority of the unsuccessful tasks for MM are due to missing deadlines, which implies energy wastage. That is why the wastage for MM is remarkably higher than \ee in low to moderate arrival rates. However, the canceled-to-missed ratio for MM gradually increases, because with the system becomes oversubscribed and the arriving tasks cannot be allocated and they are eventually dropped from the arriving queue.

\subsection{Analyzing Fairness Across Task Types}
Although \ee considers both the energy consumption and deadline constraints, it is biased towards certain task types. To address this problem, we proposed the \fee heuristic to improve the fairness across task types. %Note that task type completion rate is the ratio of the number of completed tasks on-time for a certain task type to the total number of tasks of that type arrived to the system. 
In this part, we conduct an experiment to compare the fairness of \fee against other heuristics. To this end, we utilize 30 synthesized workload traces with arrival rate of 5.0 tasks per second where each workload contains 2000 tasks. The task types and machines characteristics are described in Section~\ref{sec:exp_setup}.

\begin{figure}
     \centering
     \includegraphics[width=0.9\linewidth]{\paperfolder/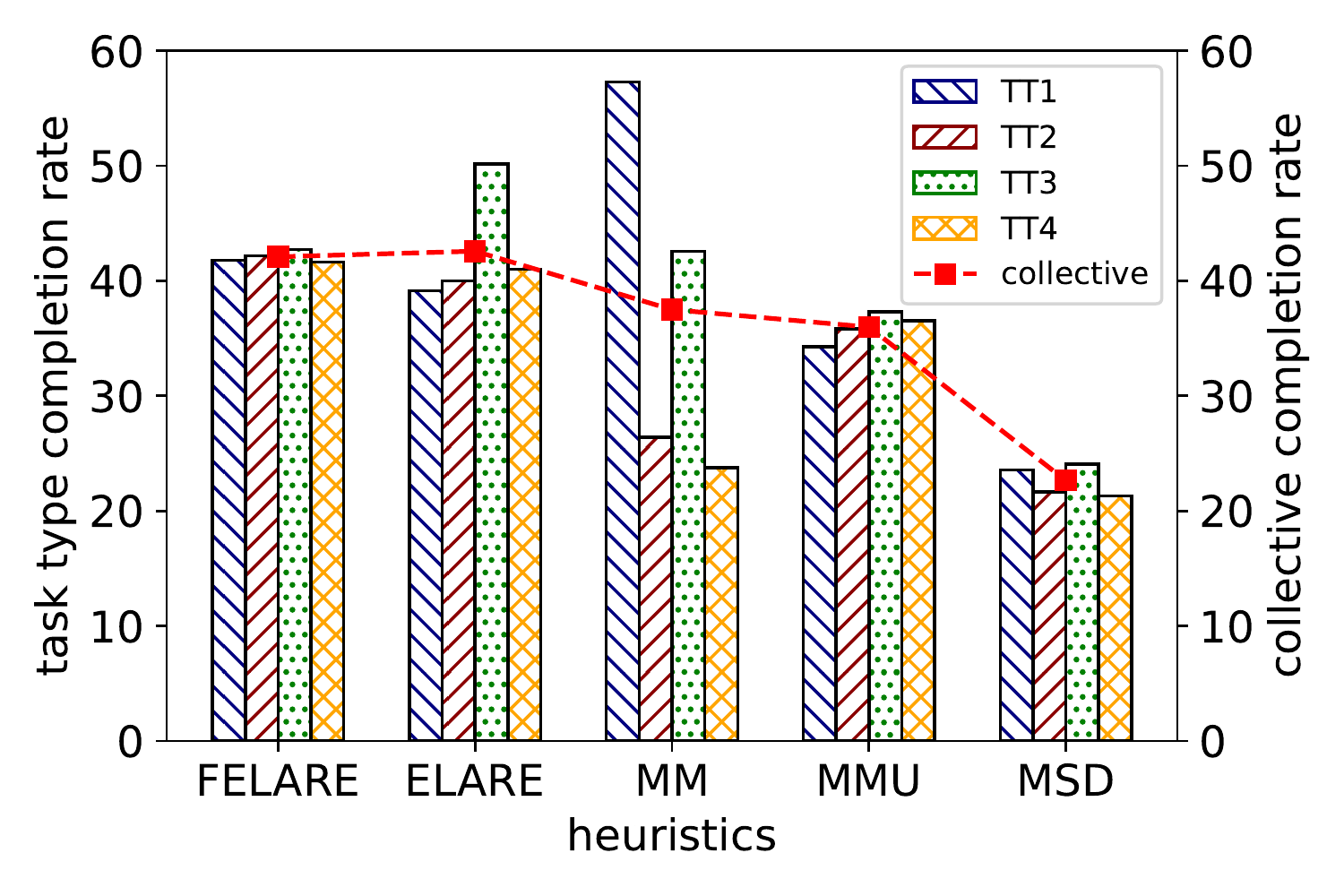}
     \vspace{-3mm}
     \caption{The fairness across task types $TT_1$ to $TT_4$ for \fee, \ee, MM, MMU, and MSD heuristics. Also, the right-side vertical axis and red data points represent the collective completion rate resulted by each heuristic.}
      \vspace{-3mm}

  \label{fig:fairness}
\end{figure}

\begin{figure}[h]
     \centering
     \includegraphics[width=0.9\linewidth]{\paperfolder/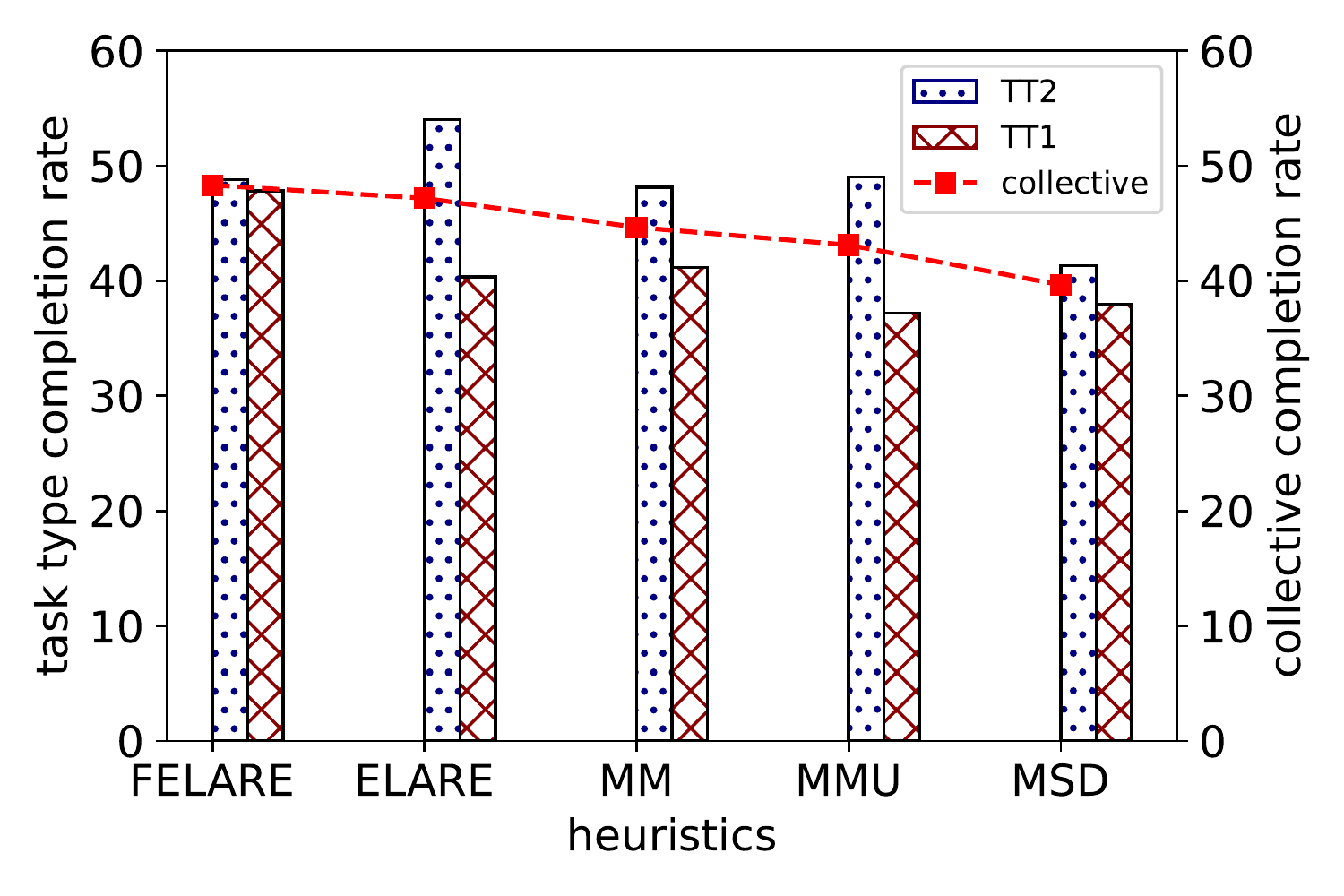}
     \vspace{-3mm}
     \caption{The figure shows the fairness across task types, face recognition and speech recognition on AWS instances for \fee, \ee, MM, MMU, and MSD. Also, the right y-axis and red curve represent the collective completion rate for each heuristic.}
      \vspace{-4mm}

  \label{fig:fairness-real}
\end{figure}

Figure~\ref{fig:fairness} shows the results of this experiment. The x-axis represents the heuristics. The left and right y-axes also represent the task type and collective completion rates, respectively. Here, collective completion rate represents the ratio of the successfully completed tasks to the total number of tasks that has arrived to the system.
We observe that \ee is biased towards $T_3$ and MM towards $T_1$ and $T_3$. However, \fee could considerably improve the fairness with negligible degradation in the total completion rate. In the case of AWS workload trace, Figure~\ref{fig:fairness-real} shows the fairness of mapping heuristics across face and speech recognition applications at an arrival rate of 2 tasks per second. The results are in agreement with the previous experiment on the synthesized workload, where the \fee method exhibits substantially higher fairness than the other heuristics. 

% Recall that \fee utilizes two methods to achieve the fairness: (i) Prioritized the suffered task types in mapping events and (ii) Drop not-suufered tasks in machine queues to make infeasible suffered tasks feasible. The degradation in total completion rate is mostly due to second mechanism in FEE. 

\section{Conclusion and Future Work}
\label{sec:conclusion}
In this research, we investigated the fair and energy-efficient allocation of concurrent and latency-sensitive ML applications in Heterogeneous Edge Computing (HEC) systems. We proposed a two-phase heuristics, called \ee, to address both energy and latency objectives. This heuristic proactively drops tasks that are unlikely to meet their deadlines, thereby, avoiding wasting energy. In the experiments, we showed that \ee could considerably reduce the wasted energy via proactively dropping infeasible tasks and choosing the machines with the minimum energy usage for each task. To address the bias to certain task types in \ee, we extended it and proposed the \fee heuristics that treats all task types fairly. \fee measures the completion rate per task type to determine the suffered task types. Then, it prioritizes the suffered task types in each mapping event. Moreover, in the case of observing infeasible suffered task types in a mapping event, \fee drops tasks from the non-suffered task types to make the infeasible suffered tasks, feasible. The evaluation results showed that devising customized mapping heuristics for the HEC systems can noticeably improve the fairness and energy consumption. 
%More specifically, the dispersion of the task type completion rates was reduced by around 47$\times$, resulting from employing \fee mapping heuristics.

In the future, we plan to extend our analysis and the heuristic for the HEC systems to the edge-to-cloud continuum. Then, the trade-off between network transfer time and the energy consumption due to local processing of the tasks needs to be investigated to eventually have a fair mapping heuristic that can fulfill both the energy and latency objectives. Another avenue for future study is to measure the heterogeneity degree of the HEC system and leverage it to dynamically apply various mapping heuristics, such that the energy and latency objectives are met.

\section*{Acknowledgement}
This research is supported by the National Science Foundation under awards\# CNS-2007209 (UL) and 2007202 (UofSC). 
\bibliographystyle{IEEEtran}
\balance
\bibliography{references}

\begin{thebibliography}{10}
\providecommand{\url}[1]{#1}
\csname url@samestyle\endcsname
\providecommand{\newblock}{\relax}
\providecommand{\bibinfo}[2]{#2}
\providecommand{\BIBentrySTDinterwordspacing}{\spaceskip=0pt\relax}
\providecommand{\BIBentryALTinterwordstretchfactor}{4}
\providecommand{\BIBentryALTinterwordspacing}{\spaceskip=\fontdimen2\font plus
\BIBentryALTinterwordstretchfactor\fontdimen3\font minus
  \fontdimen4\font\relax}
\providecommand{\BIBforeignlanguage}[2]{{%
\expandafter\ifx\csname l@#1\endcsname\relax
\typeout{** WARNING: IEEEtran.bst: No hyphenation pattern has been}%
\typeout{** loaded for the language `#1'. Using the pattern for}%
\typeout{** the default language instead.}%
\else
\language=\csname l@#1\endcsname
\fi
#2}}
\providecommand{\BIBdecl}{\relax}
\BIBdecl

\bibitem{bhatia2017comprehensive}
M.~Bhatia and S.~K. Sood, ``A comprehensive health assessment framework to
  facilitate iot-assisted smart workouts: A predictive healthcare
  perspective,'' \emph{Computers in Industry}, vol.~92, pp. 50--66, 2017.

\bibitem{salehi2016stochastic}
M.~A. Salehi, J.~Smith, A.~A. Maciejewski, H.~J. Siegel, E.~K. Chong,
  J.~Apodaca, L.~D. Brice{\~n}o, T.~Renner, V.~Shestak, J.~Ladd \emph{et~al.},
  ``Stochastic-based robust dynamic resource allocation for independent tasks
  in a heterogeneous computing system,'' \emph{{Journal of Parallel and
  Distributed Computing (JPDC)}}, vol.~97, pp. 96--111, Nov. 2016.

\bibitem{mokhtari2020autonomous}
A.~Mokhtari, C.~Denninnart, and M.~A. Salehi, ``Autonomous task dropping
  mechanism to achieve robustness in heterogeneous computing systems,'' in
  \emph{2020 IEEE International Parallel and Distributed Processing Symposium
  Workshops (IPDPSW)}.\hskip 1em plus 0.5em minus 0.4em\relax IEEE, 2020, pp.
  17--26.

\bibitem{davood22}
D.~G. Samani and M.~Amini~Salehi, ``Exploring the impact of virtualization on
  the usability of the deep learning applications,'' in \emph{Proceedings of
  the 22th IEEE/ACM International Symposium on Cluster, Cloud and Internet
  Computing}, ser. CCGrid '22, May 2022.

\bibitem{9700908}
J.~Wen, J.~Liu, F.~Xu, X.~Duan, and J.~Huang, ``Face recognition system design
  based on esp32,'' in \emph{2022 International Seminar on Computer Science and
  Engineering Technology (SCSET)}, 2022, pp. 114--116.

\bibitem{EM9D}
``Designware arc em9d / em11d processors,''
  \url{https://www.synopsys.com/dw/ipdir.php?ds=arc-em9d-em11d}, accessed:
  2022-05-24.

\bibitem{gholipour2021recent}
N.~Gholipour, E.~Arianyan, and R.~Buyya, ``Recent advances in energy efficient
  resource management techniques in cloud computing environments,'' \emph{arXiv
  preprint arXiv:2107.06005}, 2021.

\bibitem{liu2019cooper}
C.~Liu, K.~Li, J.~Liang, and K.~Li, ``Cooper-sched: A cooperative scheduling
  framework for mobile edge computing with expected deadline guarantee,''
  \emph{IEEE Transactions on Parallel and Distributed Systems}, 2019.

\bibitem{ran2018deepdecision}
X.~Ran, H.~Chen, X.~Zhu, Z.~Liu, and J.~Chen, ``Deepdecision: A mobile deep
  learning framework for edge video analytics,'' in \emph{IEEE INFOCOM
  2018-IEEE Conference on Computer Communications}.\hskip 1em plus 0.5em minus
  0.4em\relax IEEE, 2018, pp. 1421--1429.

\bibitem{huynh2017deepmon}
L.~N. Huynh, Y.~Lee, and R.~K. Balan, ``Deepmon: Mobile gpu-based deep learning
  framework for continuous vision applications,'' in \emph{Proceedings of the
  15th Annual International Conference on Mobile Systems, Applications, and
  Services}, 2017, pp. 82--95.

\bibitem{chen2020deep}
Y.~Chen, B.~Zheng, Z.~Zhang, Q.~Wang, C.~Shen, and Q.~Zhang, ``Deep learning on
  mobile and embedded devices: State-of-the-art, challenges, and future
  directions,'' \emph{ACM Computing Surveys (CSUR)}, vol.~53, no.~4, pp. 1--37,
  2020.

\bibitem{jiang2020approximate}
H.~Jiang, F.~J.~H. Santiago, H.~Mo, L.~Liu, and J.~Han, ``Approximate
  arithmetic circuits: A survey, characterization, and recent applications,''
  \emph{Proceedings of the IEEE}, vol. 108, no.~12, pp. 2108--2135, 2020.

\bibitem{mrazek2016design}
V.~Mrazek, S.~S. Sarwar, L.~Sekanina, Z.~Vasicek, and K.~Roy, ``Design of
  power-efficient approximate multipliers for approximate artificial neural
  networks,'' in \emph{Proceedings of the 35th International Conference on
  Computer-Aided Design}, 2016, pp. 1--7.

\bibitem{ansari2019improving}
M.~S. Ansari, V.~Mrazek, B.~F. Cockburn, L.~Sekanina, Z.~Vasicek, and J.~Han,
  ``Improving the accuracy and hardware efficiency of neural networks using
  approximate multipliers,'' \emph{IEEE Transactions on Very Large Scale
  Integration (VLSI) Systems}, vol.~28, no.~2, pp. 317--328, 2019.

\bibitem{venkataramani2015approximate}
S.~Venkataramani, S.~T. Chakradhar, K.~Roy, and A.~Raghunathan, ``Approximate
  computing and the quest for computing efficiency,'' in \emph{Proceedings of
  the 52nd ACM/EDAC/IEEE Design Automation Conference (DAC)}, 2015, pp. 1--6.

\bibitem{akbari2018px}
O.~Akbari, M.~Kamal, A.~Afzali-Kusha, M.~Pedram, and M.~Shafique, ``Px-cgra:
  Polymorphic approximate coarse-grained reconfigurable architecture,'' in
  \emph{Design, Automation \& Test in Europe Conference \& Exhibition
  (DATE)}.\hskip 1em plus 0.5em minus 0.4em\relax IEEE, 2018, pp. 413--418.

\bibitem{wu2020integer}
H.~Wu, P.~Judd, X.~Zhang, M.~Isaev, and P.~Micikevicius, ``Integer quantization
  for deep learning inference: Principles and empirical evaluation,''
  \emph{arXiv preprint arXiv:2004.09602}, 2020.

\bibitem{jacob2018quantization}
B.~Jacob, S.~Kligys, B.~Chen, M.~Zhu, M.~Tang, A.~Howard, H.~Adam, and
  D.~Kalenichenko, ``Quantization and training of neural networks for efficient
  integer-arithmetic-only inference,'' in \emph{Proceedings of the IEEE
  conference on computer vision and pattern recognition}, 2018, pp. 2704--2713.

\bibitem{coelho2021automatic}
C.~N. Coelho, A.~Kuusela, S.~Li, H.~Zhuang, J.~Ngadiuba, T.~K. Aarrestad,
  V.~Loncar, M.~Pierini, A.~A. Pol, and S.~Summers, ``Automatic heterogeneous
  quantization of deep neural networks for low-latency inference on the edge
  for particle detectors,'' \emph{Nature Machine Intelligence}, vol.~3, no.~8,
  pp. 675--686, 2021.

\bibitem{bitam2018fog}
S.~Bitam, S.~Zeadally, and A.~Mellouk, ``Fog computing job scheduling
  optimization based on bees swarm,'' \emph{Enterprise Information Systems},
  vol.~12, no.~4, pp. 373--397, 2018.

\bibitem{yadav2018adaptive}
R.~Yadav, W.~Zhang, O.~Kaiwartya, P.~R. Singh, I.~A. Elgendy, and Y.-C. Tian,
  ``{Adaptive energy-aware algorithms for minimizing energy consumption and SLA
  violation in cloud computing},'' \emph{IEEE Access}, vol.~6, pp.
  55\,923--55\,936, 2018.

\bibitem{zhang2018energy}
Y.~Zhang, J.~He, and S.~Guo, ``Energy-efficient dynamic task offloading for
  energy harvesting mobile cloud computing,'' in \emph{IEEE international
  conference on networking, architecture and storage (NAS)}.\hskip 1em plus
  0.5em minus 0.4em\relax IEEE, 2018, pp. 1--4.

\bibitem{ding2020q}
D.~Ding, X.~Fan, Y.~Zhao, K.~Kang, Q.~Yin, and J.~Zeng, ``Q-learning based
  dynamic task scheduling for energy-efficient cloud computing,'' \emph{Future
  Generation Computer Systems}, vol. 108, pp. 361--371, 2020.

\bibitem{zhou2019minimizing}
X.~Zhou, G.~Zhang, J.~Sun, J.~Zhou, T.~Wei, and S.~Hu, ``Minimizing cost and
  makespan for workflow scheduling in cloud using fuzzy dominance sort based
  heft,'' \emph{Future Generation Computer Systems}, vol.~93, pp. 278--289,
  2019.

\bibitem{kumar2020pso}
M.~Kumar and S.~C. Sharma, ``Pso-based novel resource scheduling technique to
  improve qos parameters in cloud computing,'' \emph{Neural Computing and
  Applications}, vol.~32, no.~16, pp. 12\,103--12\,126, 2020.

\bibitem{ghanavati2020energy}
S.~Ghanavati, J.~H. Abawajy, and D.~Izadi, ``An energy aware task scheduling
  model using ant-mating optimization in fog computing environment,''
  \emph{IEEE Transactions on Services Computing}, 2020.

\bibitem{tarafdar2021energy}
A.~Tarafdar, M.~Debnath, S.~Khatua, and R.~K. Das, ``Energy and makespan aware
  scheduling of deadline sensitive tasks in the cloud environment,''
  \emph{Journal of Grid Computing}, vol.~19, no.~2, pp. 1--25, 2021.

\bibitem{denninnart2020efficient}
C.~Denninnart, J.~Gentry, A.~Mokhtari, and M.~A. Salehi, ``Efficient task
  pruning mechanism to improve robustness of heterogeneous computing systems,''
  \emph{Journal of Parallel and Distributed Computing}, vol. 142, pp. 46--61,
  2020.

\bibitem{aazam2021task}
M.~Aazam, S.~Zeadally, and E.~F. Flushing, ``Task offloading in edge computing
  for machine learning-based smart healthcare,'' \emph{Computer Networks}, vol.
  191, p. 108019, 2021.

\bibitem{kherraf2019optimized}
N.~Kherraf, H.~A. Alameddine, S.~Sharafeddine, C.~M. Assi, and A.~Ghrayeb,
  ``Optimized provisioning of edge computing resources with heterogeneous
  workload in iot networks,'' \emph{IEEE Transactions on Network and Service
  Management}, vol.~16, no.~2, pp. 459--474, 2019.

\bibitem{rahman2021iot}
M.~A. Rahman and M.~S. Sadi, ``Iot enabled automated object recognition for the
  visually impaired,'' \emph{Computer Methods and Programs in Biomedicine
  Update}, vol.~1, p. 100015, 2021.

\bibitem{ipdps19}
J.~Gentry, C.~Denninnart, and M.~Amini~Salehi, ``Robust dynamic resource
  allocation via probabilistic task pruning in heterogeneous computing
  systems,'' in \emph{Proceedings of the 33rd IEEE International Parallel \&
  Distributed Processing Symposium}, ser. IPDPS '19, May 2019.

\bibitem{chavithcw19}
C.~Denninnart, J.~Gentry, and M.~Amini~Salehi, ``Improving robustness of
  heterogeneous serverless computing systems via probabilistic task pruning,''
  in \emph{Proceedings of the 28th IEEE International Parallel and Distributed
  Processing Symposium Workshops (IPDPSW)}, May 2019, pp. 6--15.

\bibitem{schroff2015facenet}
F.~Schroff, D.~Kalenichenko, and J.~Philbin, ``Facenet: A unified embedding for
  face recognition and clustering,'' in \emph{Proceedings of the IEEE
  conference on computer vision and pattern recognition}, 2015, pp. 815--823.

\bibitem{amodei2016deep}
D.~Amodei, S.~Ananthanarayanan, R.~Anubhai, J.~Bai, E.~Battenberg, C.~Case,
  J.~Casper, B.~Catanzaro, Q.~Cheng, G.~Chen \emph{et~al.}, ``Deep speech 2:
  End-to-end speech recognition in english and mandarin,'' in
  \emph{International conference on machine learning}.\hskip 1em plus 0.5em
  minus 0.4em\relax PMLR, 2016, pp. 173--182.

\bibitem{zhang2016joint}
K.~Zhang, Z.~Zhang, Z.~Li, and Y.~Qiao, ``Joint face detection and alignment
  using multitask cascaded convolutional networks,'' \emph{IEEE signal
  processing letters}, vol.~23, no.~10, pp. 1499--1503, 2016.

\bibitem{LFWTechUpdate}
G.~B. H.~E. Learned-Miller, ``Labeled faces in the wild: Updates and new
  reporting procedures,'' University of Massachusetts, Amherst, Tech. Rep.
  UM-CS-2014-003, May 2014.

\bibitem{ali2000representing}
S.~Ali, H.~J. Siegel, M.~Maheswaran, D.~Hensgen, S.~Ali \emph{et~al.},
  ``Representing task and machine heterogeneities for heterogeneous computing
  systems,'' \emph{Journal of Applied Science and Engineering}, vol.~3, no.~3,
  pp. 195--207, 2000.

\bibitem{praveen2018effective}
S.~P. Praveen, K.~T. Rao, and B.~Janakiramaiah, ``Effective allocation of
  resources and task scheduling in cloud environment using social group
  optimization,'' \emph{Arabian Journal for Science and Engineering}, vol.~43,
  no.~8, pp. 4265--4272, 2018.

\bibitem{shi2019performance}
Y.~Shi, Z.~Chen, W.~Quan, and M.~Wen, ``A performance study of static task
  scheduling heuristics on cloud-scale acceleration architecture,'' in
  \emph{Proceedings of the 5th International Conference on Computing and Data
  Engineering}, 2019, pp. 81--85.

\end{thebibliography}
\end{document}